\documentclass[trackchanges]{aastex63}

\usepackage{CJK}
\usepackage{graphicx}
\usepackage{enumerate}
\usepackage{amssymb, amsmath}
\usepackage{natbib}
\usepackage{color}
\usepackage{ulem}
\usepackage{dblfnote}
\usepackage{appendix}
\usepackage{hyperref}
\usepackage[stable]{footmisc}
\usepackage{comment}
\usepackage{footnote}
\usepackage{lineno}

\usepackage[nodayofweek]{datetime}
\newdateformat{monthyearday}{%
  \THEYEAR\ \monthname[\THEMONTH] \twodigit{\THEDAY}}

\newcommand{\mj}{\ensuremath{\,M_{\rm J}}}

\newcommand{\mum}{$\mu$m}
\newcommand{\epseri}{$\epsilon$ Eridani}
\newcommand{\deleri}{$\delta$ Eridani}
\newcommand{\tentos}{\ensuremath{10^{-6}}}
\newcommand{\tentoe}{\ensuremath{10^{-8}}}

\begin{document}
\accepted{AJ}
\shorttitle{Observing \epseri~with JWST/NIRCam}
\shortauthors{Llop-Sayson et al.}


\title{ Searching for Planets Orbiting $\epsilon$~Eridani  with JWST/NIRCam}

\correspondingauthor{Jorge Llop-Sayson,\\
jorge.llop.sayson@jpl.nasa.gov}

\author[0000-0002-3414-784X]{Jorge Llop-Sayson}
\affiliation{Jet Propulsion Laboratory, California Institute of Technology, Pasadena, CA 91109, USA}

\author[0000-0002-5627-5471]{Charles Beichman}
\affiliation{NASA Exoplanet Science Institute, IPAC, Pasadena, CA 91125, USA}
\affiliation{Jet Propulsion Laboratory, California Institute of Technology, Pasadena, CA 91109, USA}

\author[0000-0001-5966-837X]{Geoffrey Bryden}
\affiliation{Jet Propulsion Laboratory, California Institute of Technology, Pasadena, CA 91109, USA}

\author[0000-0001-7591-2731]{Marie Ygouf}
\affiliation{Jet Propulsion Laboratory, California Institute of Technology, Pasadena, CA 91109, USA}

\author[0000-0001-8612-3236]{Andras Gaspar}
\affiliation{Steward Observatory, University of Arizona, Tucson, AZ, 85721, USA}

\author{William Thompson}
\affiliation{NRC Herzberg Astronomy and Astrophysics, 5071 West Saanich Road, Victoria, BC, V9E 2E7, Canada}

\author[0000-0002-1838-4757]{Aniket Sanghi}
\altaffiliation{NSF Graduate Research Fellow}
\affiliation{Department of Astronomy, California Institute of Technology, Pasadena, CA 91125, USA}

\author{Dimitri Mawet}
\affiliation{Department of Astronomy, California Institute of Technology, Pasadena, CA 91125, USA}

\author[0000-0002-7162-8036]{Alexandra~Z. Greenbaum}
\affiliation{IPAC, California Institute of Technology, 1200 E. California Blvd., Pasadena, CA 91125, USA}

\author[0000-0002-0834-6140]{Jarron Leisenring}
\affiliation{Steward Observatory, University of Arizona, Tucson, AZ, 85721, USA}

\author[0000-0002-9977-8255]{Schuyler Wolff}
\affiliation{Steward Observatory, University of Arizona, Tucson, AZ, 85721, USA}

\author[0000-0002-7893-6170]{Marcia Rieke}
\affiliation{Steward Observatory, University of Arizona, Tucson, AZ, 85721, USA}

\author[0000-0003-2303-6519]{George Rieke}
\affiliation{Steward Observatory, University of Arizona, Tucson, AZ, 85721, USA}


\date{\today}

\begin{abstract}
We present observations of \epseri~with the JWST/NIRCam coronagraph aimed at imaging planets orbiting within this system. In particular, these observations targeted (1) the Jupiter-like planet, first detected orbiting at 3.5 AU with radial velocity observations, and (2) the planet postulated to be responsible for carving the edges of \epseri's outer ring, expected to orbit at 40-50 AU. However, no point sources were detected at a statistically significant level. We report new, improved upper limits at 4 $\mu$m: $\sim$1$\times$\tentos~contrast at 1\arcsec, and $\sim$2$\times$\tentoe~beyond 5\arcsec. The latter contrast limit precludes Saturn-mass planets at separations $>$16~AU given current models. We also report upper limits for \epseri's disk emission at 4 $\mu$m. While the radial surface brightness profile shows no evidence of emission, we detect a 1-$\sigma$ surface brightness signal on the east side of the system, consistent with forward scattering emission expected for \epseri's disk inclination.
Finally, we evaluate the performance of the 3-roll observation strategy, which was first employed in these observations: the gains in contrast are modest, with 20-30\% improvements with respect to the conventional 2-roll strategy.

\vspace{0.5 in}
\end{abstract}

\section{Introduction}

\epseri~is a nearby (3.2 pc), adolescent (400-800 Myr, \cite{Mamajek2008, Sahlholdt2019}) K2V star \citep{Gray2003} with K$_s$=1.75 mag. It was identified by IRAS as one of the original ``Fab Four" stars with a prominent debris disk detectable in the far-infrared ($\lambda>$25 \mum;  \citet{Gillett1986,Backman1993, Backman2009, Su2017}. The combination of IRAS, Spitzer, Herschel and SOFIA observations suggest a multi-ring structure with a ring (or possibly two rings) of warm grains situated around 2-25 AU and a colder ring of material out to 70 AU. ALMA observations find a narrow ring with a width of just 11-13 AU located 69 AU away from the star suggesting the possible presence of a planet sculpting the ring at 40-50 AU \citep{Booth2017, Booth2023}. 
The combination of 30+ years of Precision Radial Velocity (PRV) measurements has led to the secure detection of a Jupiter-like planet, \epseri~b, located at 3.5$\pm$0.04 AU \citep{Hatzes2000,Mawet2019,Llop-Sayson2021}. Recently, a study by \citet{Thompson2025} constrained the mass of this planet to 0.98$\pm$0.07 \mj, and significantly improved its orbit solutions.~
It is within this context that the Guaranteed Time Observers (GTO) have planned a series of observations taking advantage of multiple JWST instruments and modes. 

This program (\#1193) includes both NIRCam and MIRI observations, the latter focused on the properties of the well-known debris disk. This paper concentrates exclusively on the NIRCam results whose aim is to find  point sources which might be planets in the \epseri~system. The MIRI results will be presented by Wolff et al. (in preparation).

Section $\S$\ref{sec:observations} describes the observations and $\S$\ref{sec:reduce} the data reduction steps. $\S$\ref{sec:results} presents the results from our observations: PSF-subtracted images close to the star and over a wider field and the new limiting contrast levels ($\S$\ref{sec:results_sensitivity}), the search for scattered disk emission ($\S$\ref{sec:results_scattered}), and the list of sources detected in the images ($\S$\ref{sec:results_sources}).  $\S$\ref{sec:discuss} discusses the detection/limits for a sub-Jovian mass planet  in the context of the dynamical detection of such a planet, as well as the three roll observation strategy performance.

\section{Observations}\label{sec:observations}

\begin{deluxetable*}{llll}[t!]
\tabletypesize{\scriptsize}
\tablewidth{0pt}
\tablecaption{Properties of the Host Star \epseri\label{tab:star}
}
\tablehead{
\colhead{Property} & \colhead{Value}& \colhead{Units}& \colhead{Comments} }
\startdata
Spectral Type & K2V & &\\
T$_{\rm eff}$ &5020& K & \citet{Rosenthal2021} \\
Mass &0.82 & M$_\odot$&\citet{Rosenthal2021}\\
Luminosity & 0.33& L$_\odot$&\\
Age &400-800 &Myr & \citet{Llop-Sayson2021}\\
$[$Fe/H$]$ &$-$0.044& dex &\citet{Rosenthal2021}\\
log(g)&4.59$\pm$0.014&cgs &\citet{Rosenthal2021}\\	 
Planet inclination ($i$)&78.8\arcdeg $^{+29}_{-22}$&&\citet{Llop-Sayson2021}\\	
Disk inclination ($i$)&34$\pm$2\arcdeg&&\citep{Booth2017}\\	
R.A.\ (Eq 2000; Ep 2000)&\multicolumn{2}{l}{03$^h$32$^m$55.844$^s$\ \ $-$09\arcdeg27$^\prime$29.739\arcsec} & Gaia DR3\\
R.A.$^*$\ (Eq 2000; Ep 2023.5989)&\multicolumn{2}{l}{03$^h$32$^m$54.310$^s$\ \ $-$09\arcdeg27$^\prime$29.152\arcsec}  & \\
R.A.$^*$\ (Eq 2000; Ep 2024.0931)&\multicolumn{2}{c}{03$^h$32$^m$54.237$^s$\ \ $-$09\arcdeg27$^\prime$29.327\arcsec} & \\
Distance & 3.22$\pm$0.001 & pc & Gaia DR3\\
Proper Motion ($\mu_\alpha,\mu_\delta$)&(-974.758, 20.876) &mas/yr &Gaia DR3\\
V & 3.73 mag (121 Jy) & &\citet{Johnson1966}\\
K & 1.67 mag (143 Jy) & &\citet{Johnson1966}\\
F210M & 1.67 mag (148 Jy TBD) & &See text\\
F444W & 1.64 mag (40.6 Jy TBD) & &See text\\
\enddata
\tablecomments{$^*$As observed from vantage point of JWST L2 orbit at the given epoch.}
\end{deluxetable*}

We observed \epseri~ with JWST/NIRCam  simultaneously at two wavelength bands, F210M and F444W, one each in the long and short wavelength arms of the instrument  (\citealt{Rieke2023}; Table~\ref{tab:exposures}), and using the round M335R coronagraphic mask. 
We  first observed the star  on 2023-08-08  using the full-array mode at two roll angles to cover both the region close to the star and also approximately half the area subtended by the disk seen at  ALMA wavelengths  \citep{Booth2017}. We subsequently observed the star on 2024-02-04\&05 in subarray mode with three roll angles to  optimize the contrast close to the star and  in full-array mode with two rolls to capture the full extent to the ALMA disk.  

The exposure time at F444W was chosen to search for planets down to $<$ 1 \mj\  masses at 4\arcsec\ assuming a 5 nm wavefront drift and using representative models for giant planets, e.g. \citet{Spiegel2012}. At this separation, we should be able to detect a 1 \mj\ planet with a SNR of about 5 for an assumed age of 400-800 Myr \citep{Mamajek2008,Sahlholdt2019}. The simultaneous F210M observations were  made  to a depth adequate to identify and reject  (in a preliminary manner) background stars or extragalactic objects based on their [F210M]-[F444W] color. 

We adopted the star $\delta$ Eri (a K1 IV star \citep{Gray2003} with K=1.43 mag, compared with K=1.67 mag for \epseri) as a Point Spread Function (PSF)  reference. The star is at a separation of 2.56\arcdeg  on the sky; for the observing dates in question, the changes in solar offset angle between the two stars were  $\sim$2.3\arcdeg. The small angular displacement minimizes thermal drifts in the telescope \citep{Perrin2018}. The colors of the two stars are very similar ($\Delta$(V-K)=-0.11 mag and $\Delta$(K-[IRAS 12])= 0.015 mag in the sense \epseri$-$~$\delta$ Eri). Their $T_{eff}$'s are also very similar, 5145K for $\epsilon$ Eri and 5182K for $\delta$ Eri \citep{Bermejo2013}. According to the JWST documentation, \citep{JDoxColor}, the minor color difference will have very little effect on the achievable contrast.  

Table~\ref{tab:exposures} describes the observing parameters for the NIRCam program. We used the 5-POINT dither pattern on \deleri~to increase the diversity in the PSF for post-processing and thus to increase the contrast gain at close separations. 
Although each image in the  5-point dither for $\delta$ Eridani had shorter integration times  than  the  \epseri\  images, the reference star received approximately 1.5-2 times more  total integration time.

\begin{deluxetable*}{llllccrl}
\tabletypesize{\scriptsize}
\tablewidth{0pt}
\tablecaption{NIRCam Imaging Parameters (PID:\#1193)\label{tab:exposures}
}
\tablehead{
\colhead{Obs ID}&\colhead{Roll}&\colhead{Star}&\colhead{Filter}&\colhead{\# Group}&\colhead{\# Ints}&\colhead{Exp Time}&\colhead{Date Obs}}
\startdata
\multicolumn{8}{l}{\textit{Subarray Imaging (SUB320)}}\\
51&1 ($-$5\arcdeg)&\epseri&F210M \& F444W&10&122&2,741&2024-02-05\\
52&2 (0\arcdeg)&\epseri&F210M \& F444W&10&122&2,741&2024-02-05\\
53& 3 ($+$5\arcdeg)&\epseri&F210M \& F444W&10&122&2,741&2024-02-05\\
55&NA&$\delta$ Eri&F210M \& F444W&10&40&899$\times$5$^*$&2024-02-05\\
\multicolumn{8}{l}{\textit{full-array Imaging}}\\
61&1 ($-$5\arcdeg)&\epseri&F210M \& F444W&3&40&1,707&2023-08-08\\
63&2 ($+$5\arcdeg)&\epseri&F210M \& F444W&3&40&1,707&2023-08-08\\
64&NA&$\delta$ Eri&F210M \& F444W&3&12&505$\times$5$^*$&2023-08-08\\
50&1 ($-$5\arcdeg)&\epseri&F210M \& F444W&3&40&1,707&2024-02-04\\
54&3 ($+$5\arcdeg)&\epseri&F210M \& F444W&3&40&1,707&2024-02-05\\
56&NA&$\delta$ Eri&F210M \& F444W&10&12&505$\times$5$^{*}$&2024-02-05\\
\enddata
\tablecomments{All Observations obtained in RAPID readout mode. $^{*}$Observed at five dither positions.
}
\end{deluxetable*}

\section{Data Reduction and Post Processing\label{sec:reduce}}

The pipeline processing and post-processing steps closely follow the procedures described in \citet{Ygouf2024, Beichman2024} and are summarized below.

\subsection{Image Calibration} 

All the images shown in Table~\ref{tab:exposures}
were processed using the {\tt JWST} pipeline version \rm{2023\_3b}, calibration version \rm{1.9.6}, CRDS context for reference files \rm{jwst\_1202.pmap}, 
photometry reference file \rm{jwst\_nircam\_photom\_0157.fits}, and
distortion reference file \rm{jwst\_nircam\_distortion\_0173.asdf}. 
The dataset can be obtained at: \dataset[https://doi.org/10.17909/6dgc-5086]{https://doi.org/10.17909/6dgc-5086}.

The standard JWST pipeline \citep{jwst2022} was used with some modifications: (1) dark current corrections were ignored during processing; (2) the ramp fitting step was done by utilizing features from the {\tt SpaceKLIP} package \citep{Kammerer2022}, this reduced the 1/f noise and significantly improved the noise floor in the subarray images; and (3) measurements with only a single group before saturation were accepted to reduce saturation effects in the full-array images.

\subsubsection{Bad Pixel Rejection}

The bad pixel flagging was less conservative than the default: {\tt n\_pix\_grow\_sat} was set to 0, rather than 1. To identify truly bad pixels, we used the pipeline {\tt DQ} flags so that any pixels flagged as {\tt DO\_NOT\_USE}, e.g.\ dead pixels, those without a linearity correction, etc., were set to {\tt NaN}. 5-$\sigma$ outliers -- temporally within sub-exposures or spatially within a 5x5 box -- were also rejected. Additional bad pixels which became apparent following PSF subtraction (\S\ref{PSFsubtraction}) were similarly rejected.

\subsection{Image Registration}\label{sec:image-registration}
To estimate the position of the star behind the coronagraph mask, we performed a cross-correlation of a model PSF with the data. All simulated PSFs were computed with \texttt{STPSF} \citep{perrin14} and \texttt{WebbPSF\_ext} \citep{Leisenring2024}. Similar to \citet{Greenbaum2023}, we used the function \texttt{chi2\_shift} from the \texttt{image-registration} Python package\footnote{https://image-registration.readthedocs.io/} to compute this estimation. Our measured centroiding error of $\sim$7 mas is consistent with that of \citet{Carter2023}.

\begin{figure*}[t!]
\centering
\includegraphics[width=0.9\textwidth]{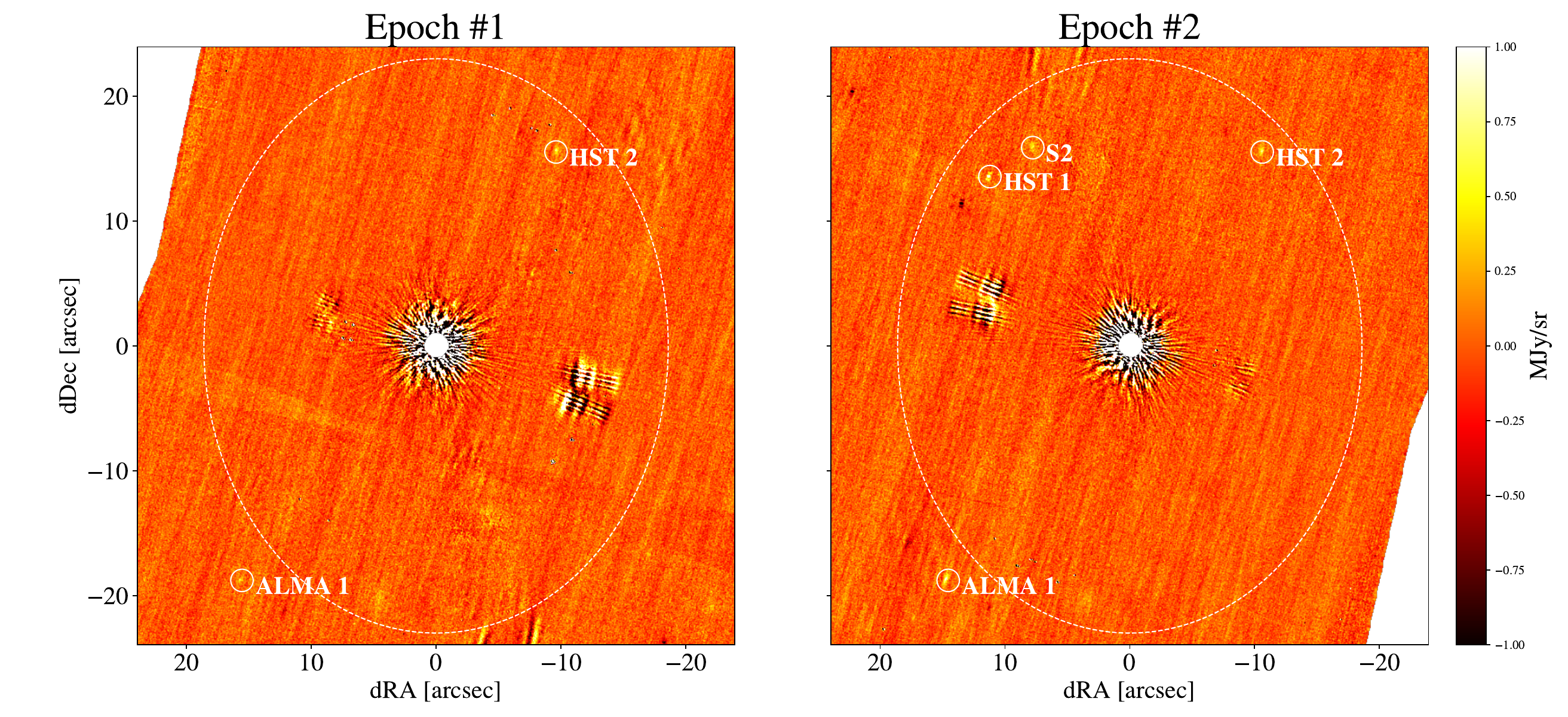}
\caption{PSF-subtracted images of the observations of \epseri~observations with the JWST/NIRCam coronagraph. Two epochs were observed separated by six months to capture the whole disk. Here shown are the images from the full-array mode observations, which were optimized to be sensitive low-mass planets beyond 2-3\arcsec. Sources HST~1 and 2 were previously detected in HST observations \citep{Wolff2023}, and ALMA~1 by ALMA~\citep{Booth2023}; source S2 is the only previously undetected source. However, it exhibits an elongated structure indicating that it is likely a background galaxy. Sources S2 and HST~1 disappear during Epoch \#1 (\textit{left}) because they were blocked by the coronagraph optical mount at the edge of the NIRCam detector. The white circle indicated the position of the ALMA disk \citep{Booth2023}. {The strong noise features appearing east and west are resulting from bright diffraction features associated with the coronagraphic substrate and its mounting plate.}
\label{fig:NIRCamSources}}
\end{figure*}

\subsection{Point Spread Function (PSF) Subtraction}\label{PSFsubtraction}
To look for faint companions orbiting \epseri, we performed PSF subtraction to reduce  the residual starlight speckles that the coronagraph could not fully suppress. In Reference Differential Imaging (RDI), the $\delta$ Eri PSF was used to perform this subtraction. Additionally, the images from the \epseri~observation rolls were utilized as references in Angular Differential Imaging (ADI). We performed the whole PSF subtraction process with the open source Python package \texttt{pyKLIP} \citep{Wang2015}. To minimize the speckle noise we used a PCA-based algorithm \citep{Amara2012} via a Karhunen Lo\'eve Image Projection (KLIP; \citet{Soummer2012}) using both the reference frames and the science frames with the different roll angles in the PSF library. Two regions were treated differently during this process given the level of residual starlight: the inner, speckle-dominated region, and the outer, background-dominated region.

\subsubsection{Outer Region}
In the outer region, typically beyond 2\arcsec~for NIRCam's M335R coronagraph, the background is the dominant source of noise. Therefore, a less aggressive use of KLIP is usually warranted to avoid the risk of over-subtracting astrophysical signals. The RDI+ADI strategy was used to remove excessive speckles and the large diffraction spikes from the NIRCam coronagraph PSF. The full-array images were processed without dividing them into annuli. 

\subsubsection{Inner Region}
In the inner, speckle-dominated region, the residual speckles from \epseri~are the dominant source of noise even after PSF subtraction. For NIRCam's M335R coronagraph this region typically extends to $\sim$2\arcsec. Given that the full-array frames have not been optimized for small angular separations, they have significant saturation; therefore, only the subarray images were utilized for this analysis. We used 7 principal components for the KLIP subtraction to take full advantage of the diversity of the dithered reference images and the two additional rolls. Each science image was divided radially into annuli, the width of which was selected to be$\sim$2.6$\times$FWHM. Each annulus was further divided into 4 azimuthal sections.

\subsection{Post-processing Optimized for Extended Emission\label{sec:diskproc}}


The F444W images, both the FULL and SUBARRAY integrations, were reprocessed using classical reference
differential imaging methods to increase the signal-to-noise (SNR) ratio beyond 2$^{\prime\prime}$. 
We chose RDI over KLIP or NMF, as it has been proven to be the most effective at retrieving faint 
extended sources of emission for both ground-based and JWST high contrast imaging data
\citep{Lawson2022,Lawson2024}. 

For this processing, we manually determined the offsets between each target-PSF pairing and the appropriate PSF scaling, by examining their subtraction residuals by eye. {We found this method to be the most efficient, likely because at separations greater than 2\arcsec, but still relatively close to the star, sparse and noisy diffraction speckles interfere with other automated approaches.}
Centering of the target images was determined for this post-processing routine by minimizing residuals between them and their 180$^{\circ}$ rotated versions, which again places emphasis on the outer regions.
Offsets and scalings were determined using the standard pipeline-processed and stage 2 median-combined 
integrations. The PSF subtracted FULL images were corrected for the quadrant bias offsets and all images
were corrected for residual 1/f noise patterns (horizontal stripes). Individual image masks were 
generated to block residual subtraction artifacts as well as the generic coronagraphic substrate and its
mounting plate. We used pre-launch high SNR cryovac testing data aligned to the location of post-launch 
flat field images taken in program 1063 to produce this mask. The residual noise of each individual 
subtracted image depends greatly on the relative offset between the target and reference source
behind the coronagraphic mask. Small offsets result in clear, mostly residual free, subtractions while
even minor offsets will produce heavy striation patterns.

\subsection{Contrast Calibration\label{sec:contrast_calibration}}
To compute the contrast limits we normalized the flux to a synthetic peak flux.  We computed  \epseri's flux density in the NIRCam bands by convolving a Kurucz model of 5020 K and \textit{log g}=4.59 with the JWST bandpasses to obtain 148 Jy in the F210M filter and 40.6 Jy at F444W assuming V=3.73 mag \citep{Johnson1966}. To estimate the flux at the peak of the off-axis PSF we simulated a set of PSFs using \texttt{STPSF} \citep{perrin14}. Measured fluxes in the NIRCam images were divided by these estimated stellar fluxes to obtain contrast ratios.

\section{Results\label{sec:results}}

\subsection{Sensitivity Limits}\label{sec:results_sensitivity}
We treat the outer, background-dominated region, where we look for large separation planets, independently from the inner, speckle-dominated region, in which we search for \epseri's Jupiter analog, planet b.

\subsubsection{Outer Region ($>$6 AU)}
The reduced images from the full-array mode observations are shown in Fig.~\ref{fig:NIRCamSources}. A handful of sources were detected within or close to the ALMA ring, none of which exhibit signs of being a bound point-source. We present details of these sources in Sec.~\ref{sec:results_sources}. 

\begin{figure}
\centering
 \includegraphics[width=0.48\textwidth]{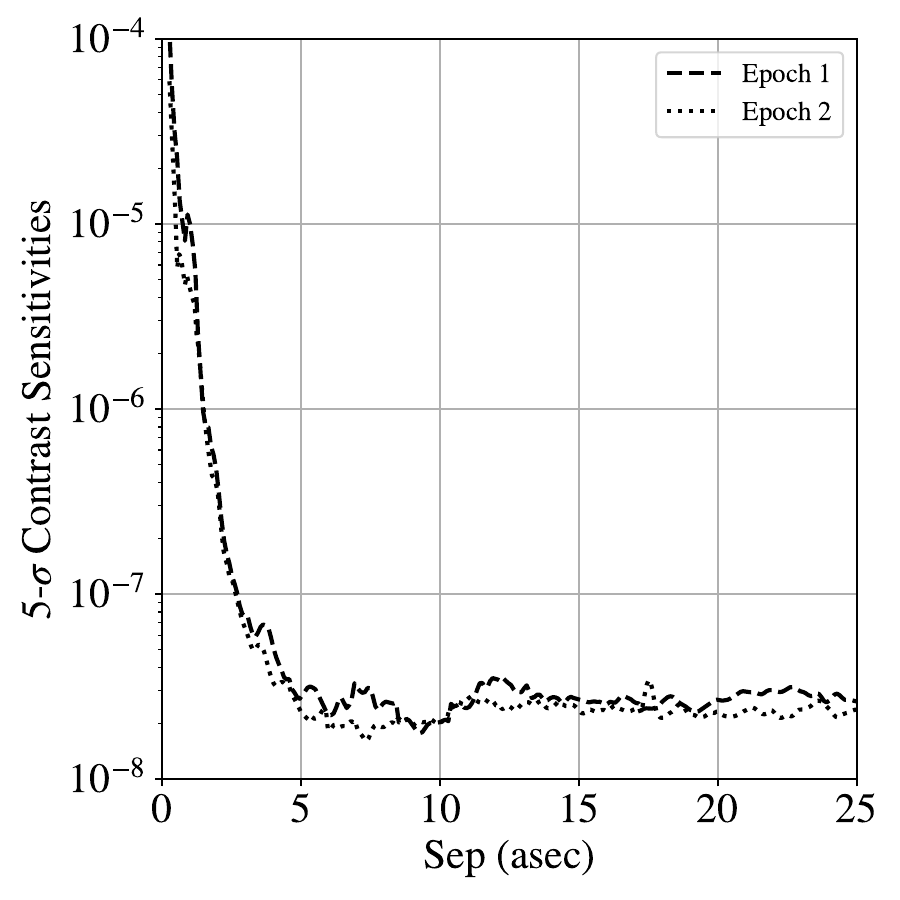} 
\caption{Full-array observation contrast curves. The exposure time for this mode was optimized for large angular separations (beyond 1.5\arcsec); the subarray mode contrast curve in Fig.~\ref{fig:sub_contours} shows the sensitivity in the inner region.
\label{fig:contrast_curve_FA}}
\end{figure}

The 5-$\sigma$ contrast curves shown in Figure~\ref{fig:contrast_curve_FA} are obtained using \texttt{pyKLIP}, with the contrast calibration done as described in Sec.~\ref{sec:contrast_calibration}. The process to compute these contrast limits can be found in detail in \citep{Greenbaum2023,Ygouf2024}. The throughput from the mask is accounted for via modeling of the instrument with \texttt{STPSF}, the algorthmic throughput is corrected via injection-recovery of fake sources, and the contrast values are adjusted to account for small sample statistics as per \citet{Mawet2014}. 

Given the deep levels of contrast achieved beyond 5\arcsec, $\sim$2$\times$\tentoe, mass estimates are challenging to compute. Evolutionary models typically work for temperatures $>$150~K, as is the case of \citet{Linder2019}. {According to their models, a contrast level of 2$\times$\tentoe~would be explained by a Neptune-sized planet. Therefore}, we can rule out young, T$_{eff}>$150K, planets {that are at least smaller than Saturn} at separations $>$5\arcsec, or 16 AU{; but o}ur lack of theoretical knowledge for this regime of low flux prevents us from giving more specific mass limits at these separations.

\subsubsection{Inner Region ($<$6 AU)}\label{sec:results_inner_region}
\begin{figure}
\centering
 \includegraphics[width=0.45\textwidth]{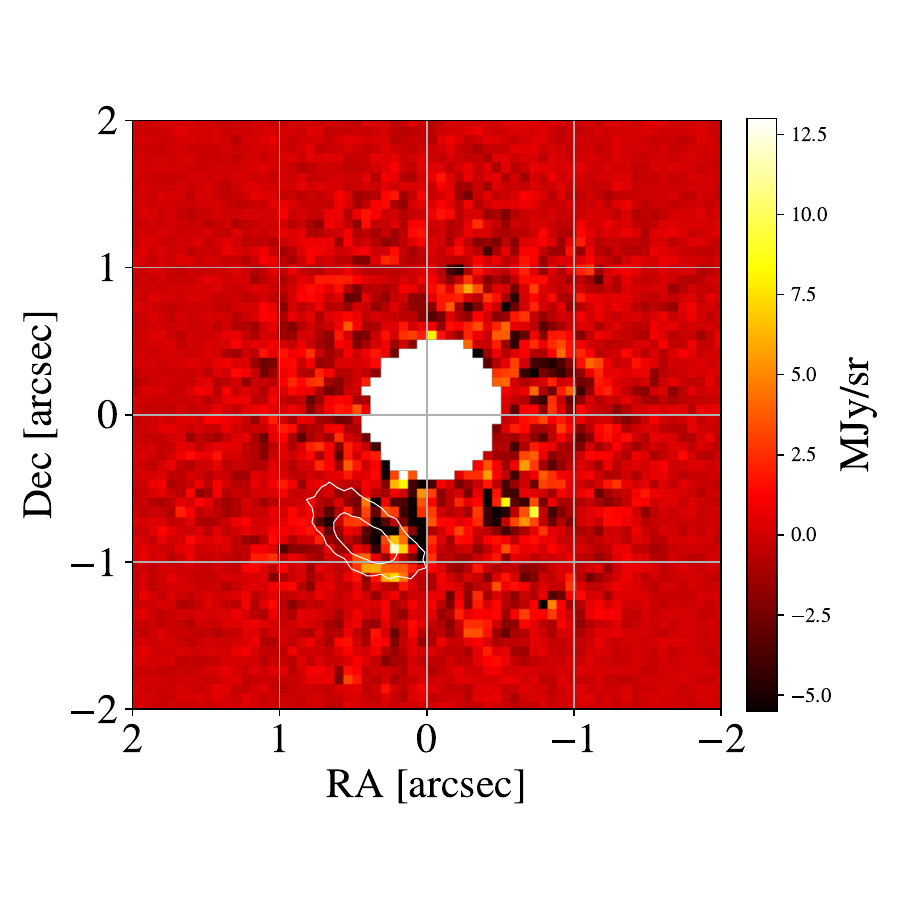} 
\caption{PSF-subtracted image of the inner region. The white lines represent the predicted location contours (1- and 2-$\sigma$) of planet b by \citet{Thompson2025} during the time of the JWST observations, Feb 4, 2024. A blob within the expected position persists in PSF subtraction; however, given its proximity to the brightest speckle in the raw data and the noise structure in that location, we cannot rule out it being a residual from data reduction.
\label{fig:sub_contours}}
\end{figure}

The PSF-subtracted image for the subarray mode in epoch \#1 is shown in Fig.~\ref{fig:sub_contours}. The 1- and 2-$\sigma$ position contours for planet b from \citet{Thompson2025} are overplotted in white. A feature consistent with a point-source stands out in the final image within the expected position, and is recovered with independent reduction processes. Its flux is consistent with the latest mass estimate for planet b given by \citet{Thompson2025}. However, its position is within 1~$\lambda/$D from one of the bright, six-fold speckles from the NIRCam coronagraph PSF (we refer to these as hexpeckles given their distribution and origin). There are three particularly bright hexpeckles in the raw data that leave an increased correlated noise at their locations in the final image: at PAs $\sim$160, $\sim$220 and $\sim$280. Our coronagraph PSF modeling suggests that the increased brightness of some hexpeckles with respect to the others is due to a pupil shift. A pupil shift in x and y of -1.5 and 0.8\% of the pupil, respectively, results in a good match (under a least-squares metric) of a \texttt{STPSF} model with respect to the raw images. The large brightness differences near these hexpeckles pose challenges with interpolation when rotating and shifting the images. The resulting noise at this location, and the other two locations with similarly bright hexpeckles, is structured and non-Gaussian. We thus conclude that there is not sufficient confidence in this feature being of astrophysical nature to report a detection. For completeness, however, we include it, labeled as S~1, in Tab.~\ref{tab:JWSTsources}, and discuss it Sec.~\ref{sec:planetb}.

\begin{figure}[t!]%
\centering
 \includegraphics[width=0.48\textwidth]{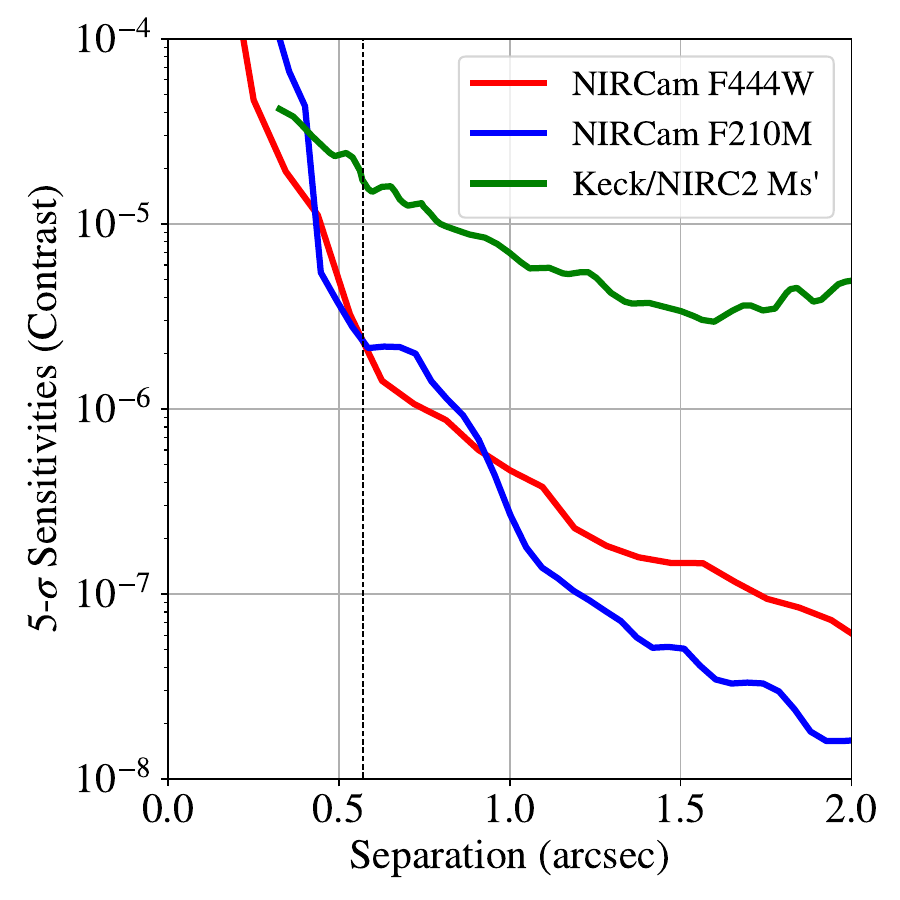} 
\caption{Sensitivity limits (5-$\sigma$) in terms of contrast. The result from these observations with JWST/NIRCam (red and blue) are compared with \citet{Llop-Sayson2021}~Keck/NIRC2 vortex data. {The F444W curve does not include the regions with highly correlated noise associated with the three bright \textit{hexpeckles} (see Fig.~\ref{fig:sub_contours}).} The dashed vertical line indicates the inner working angle of the coronagraph.
\label{fig:contrastlimits}}
\end{figure}

In Fig.~\ref{fig:contrastlimits} we report the contrast curves for the subarray observations. These contrast sensitivities are computed in the same way as described in the previous section, the difference being that the regions in the F444W data with highly correlated noise due to the three bright hexpeckles were masked out. {The noise in these three zones is not speckle noise and is non-Gaussian so we do not report a detection limit in this case.} 

\begin{figure}[t!]%
\centering
 \includegraphics[width=0.48\textwidth]{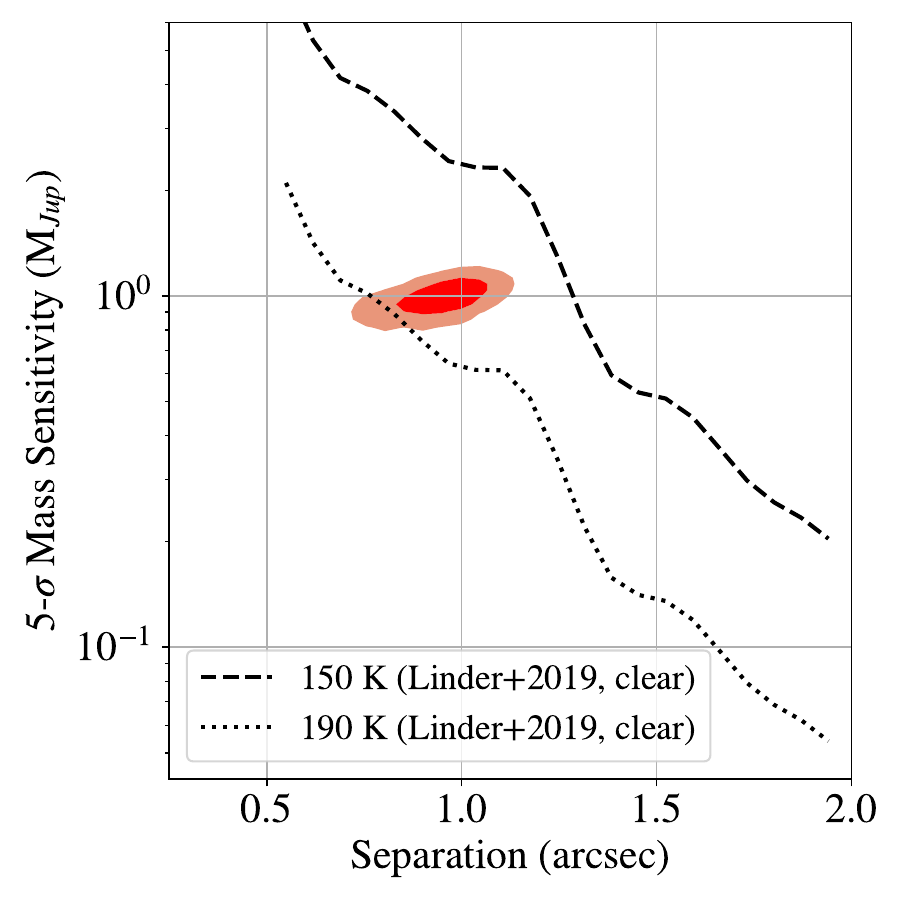} 
\caption{ Mass sensitivities from JWST/NIRCam coronagraphic observations of \epseri~assuming \citet{Linder2019} model grids. We assume a Jupiter-like planet with log(g)=3.39, and metallicity equal to the star. Red contours indicate mass estimates for planet b from \citet{Thompson2025}. A significant part of these contours contain the structured noise resulting from a bright \textit{hexpeckles} (see Fig.~\ref{fig:sub_contours}), which is not accounted for in these curves. The two temperature bounds are chosen according to \citet{Linder2019} cooling curves shown in their Fig.~6.  
\label{fig:mass_sensitivities}}
\end{figure}

From the F444W contrast curves we derive mass sensitivities with models from \citet{Linder2019}. We show mass sensitivities in Fig.~\ref{fig:mass_sensitivities} assuming a Jupiter-analog planet, i.e. log(g) of 3.39, and a metallicity equal to that of \epseri: [Fe/H]=-0.08. The mass posterior distribution from \citet{Thompson2025}'s planet b model is overplotted in red. 
Since the areas near the three brightest hexpeckles were masked out due to the highly correlated noise, the completeness implied by the mass contours of b and the mass sensitivities does not include these three regions. Moreover, as seen in Fig.~\ref{fig:sub_contours} the expected position of b lies near the brightest hexpeckle.

\begin{figure*}[t!]%
\includegraphics[width=0.48\textwidth]{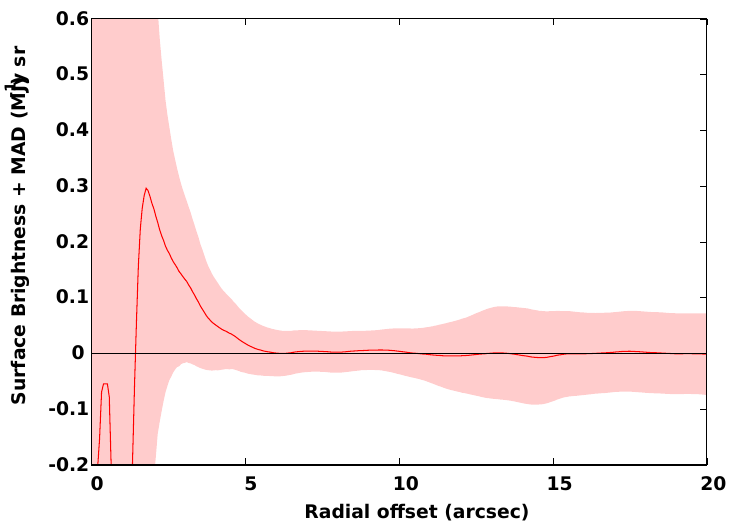} 
\includegraphics[width=0.48\textwidth]{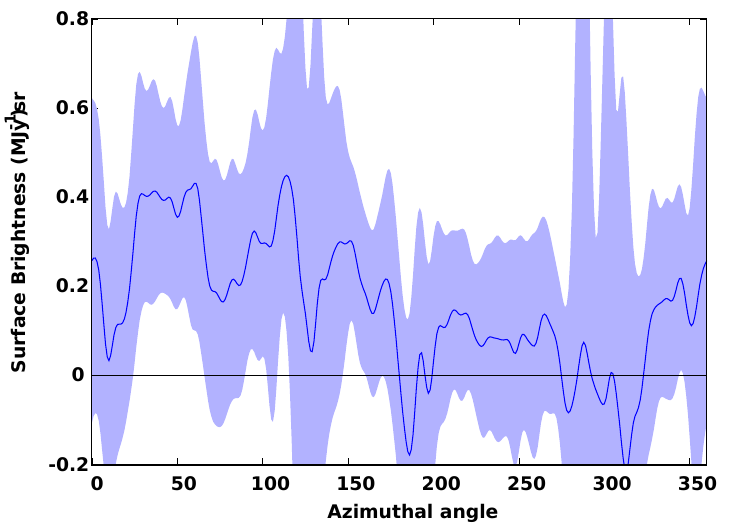} 
\caption{\textit{Left:} Azimuthal average of the surface brightness as a function of 
offset from $\epsilon$ Eridani, using the images optimized for low surface brightness 
extended emission recovery. \textit{Right:} The surface brightness as a function of
azimuth angle for between radial distances of 2\farcs2 and 3\farcs5. The East side of 
the inner disk region has an excess surface brightness of over 1$\sigma$ standard deviation
above the background. This is the forward facing side of the system, therefore larger 
scattered light emission would be expected here from an extended inner disk. The E-W asymmetry
is also apparent in the image (Figure \ref{fig:deep}).
\label{fig:azimuth}}
\end{figure*}

\subsection{Limits to Scattered Disk Light }\label{sec:results_scattered}
\begin{figure*}
\centering
\includegraphics[width=0.9\textwidth]{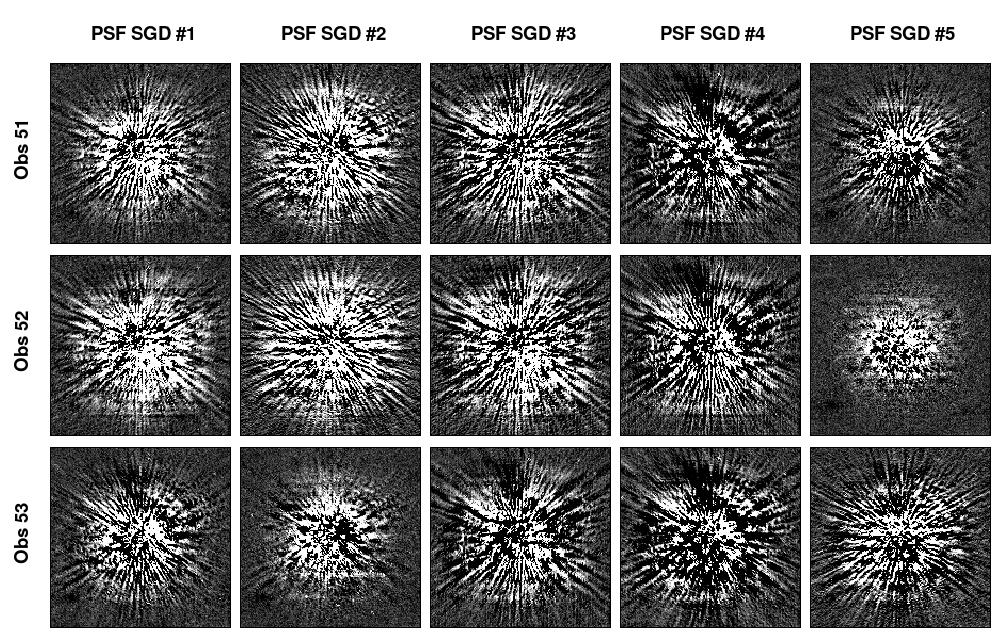}
\caption{Subtraction residuals for each target-PSF pairing used in the high SNR reduction (classic RDI) optimized for the extended emission reduction. Notably, the SGD position \#5 is very precisely aligned with Observation \#52, resulting in a nearly perfect, residual free subtraction. Images (170$\times$170 px) 
are scaled linearly between -1 and +2 MJy sr$^{-1}$.
\label{fig:residuals}}
\end{figure*}

\begin{figure*}
\centering
\includegraphics[width=0.98\textwidth]{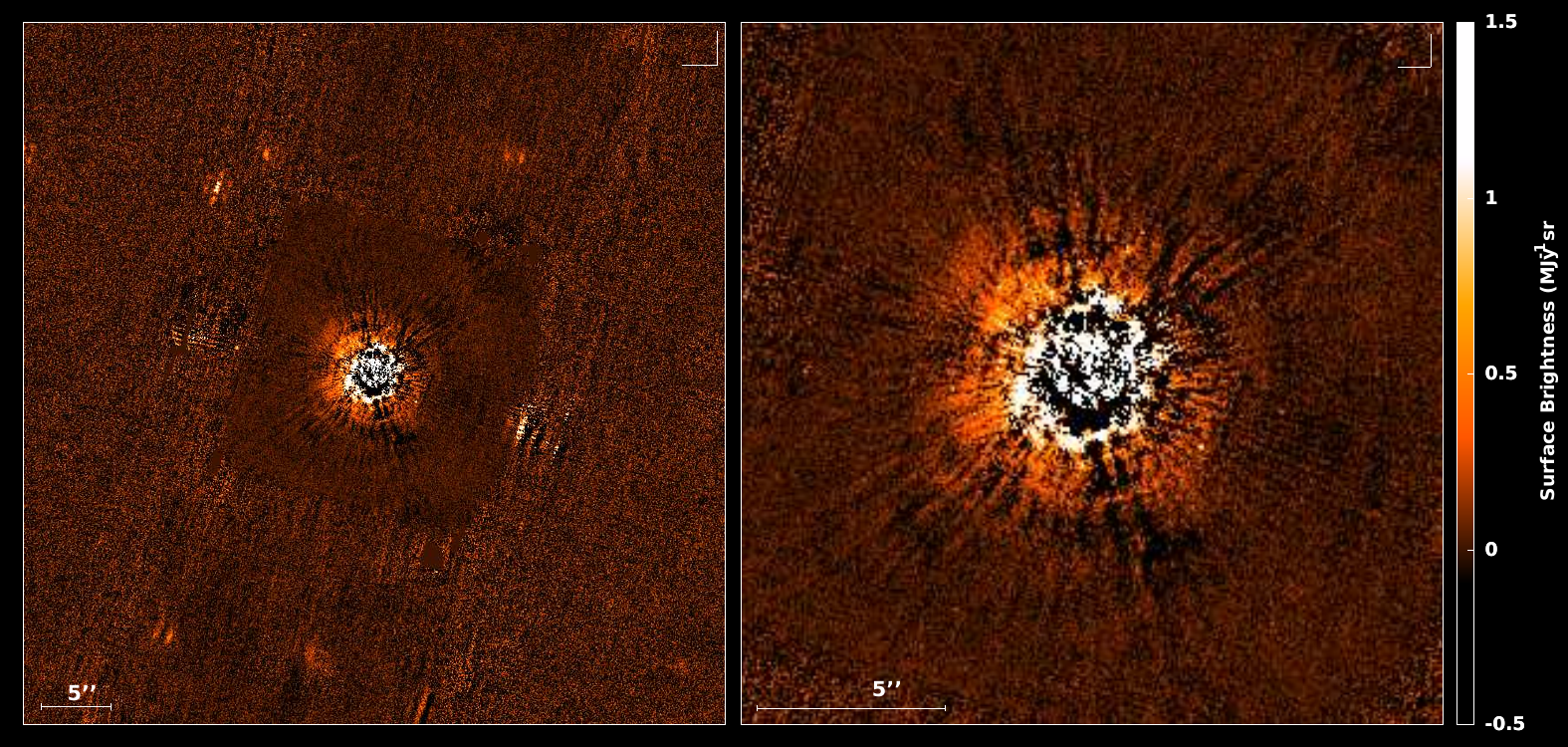}
\caption{The final F444W image produced via classical RDI optimized for extended source emission retrieval (FULL and Subarray combined), shown at a larger field of view to include the outer ring in the left hand panel 
and zoomed in to the inner regions in the right panel. While the outer ring is not detected in our observations, the inner regions  exhibit a marginal (1 $\sigma$), low level of asymmetric extended emission. The background sources are ``doubled'' where the FULL images overlap that were taken 6 months apart).
\label{fig:deep}}
\end{figure*}

\begin{figure}
\centering
\includegraphics[width=0.48\textwidth]{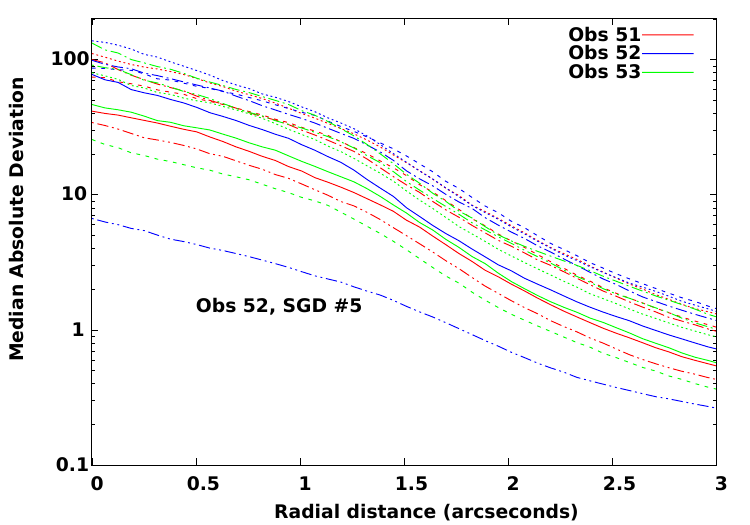}
\caption{The radial Median Absolute Deviation (MAD) values of the individual target-SGD PSF pair subtractions for the epoch 2 subarray observations (shown in Figure \ref{fig:residuals}),
used as weights for the final image combination produced for the post-processing optimized for extended emission. The SGD position \#5 subtraction for Obs \#52 is highlighted, being a remarkably well aligned pair
resulting in minimal residuals.
\label{fig:mads}}
\end{figure}

In Figure \ref{fig:residuals}, we show the
subtraction residuals in the reduction sequence described in Sec.~\ref{sec:diskproc} for the $\epsilon$ Eridani F444W subarray observations (Observation 51, 52, and 53) for each of the five sub-grid dither (SGD) pattern positions. The PSF
observed at dither position \#5 is obviously well matched for the target observed in Observation 52, as there are no apparent residual striations present. To achieve the highest SNR possible at all working angles, we combine these subtracted images by weighting their contributions based on their radial median absolute deviations (MAD). In Figure \ref{fig:mads}, we show the MAD curves of the residual images
shown in Figure \ref{fig:residuals}. The final F444W image, optimized for the detection of extended features, is shown in Figure \ref{fig:deep}, combining both the FULL (both epochs) and Subarray (epoch 2) observations.

The full and subarray observations from the two epochs were combined as described above ($\S$\ref{sec:diskproc}). The radial surface brightness profile of the final product,  smoothed with a $\sigma=5~{\rm px}$ gaussian, is shown in the left panel of
Figure \ref{fig:azimuth}. Outside of $\sim$5\arcsec\ the F444W surface brightness 
is consistent with $0\pm0.05$ MJy sr$^{-1}$. Interior to $\sim$5\arcsec there is a 
rise to $0.3$ MJy sr$^{-1}$ but with a large uncertainty, $0.3\pm0.3$ MJy sr$^{-1}$. 
We also examine the surface brightness within 2\farcs2 and 3\farcs5 as a function of 
the azimuthal angle in the right panel of Figure \ref{fig:azimuth}. While 
there is no evidence for significant emission from the disk in the radial surface
brightness profile, the east side of the system does have a surface brightness that
is 1$\sigma$ above zero, while the west side shows no significant  excess emission.
This is the type of forward scattering emission we would expect from a moderately
inclined disk. 

Our measured surface brightness values can be compared with visible light (0.6 $\mu
$m) HST/STIS observations \citep{Krishnanth2024}, where the best reduction method
yields limits of a few  tens of  MJy sr$^{-1}$ at separations of 2\arcsec$\sim$3\arcsec\
(Table~\ref{tab:disk}). Extrapolation of the visible limits must  account for the
reduced flux density of the star between the HST observation to JWST data 
at 4.44 $\mu$m, a factor of  0.33 for a K2V star ($V-K_s$=2.0 mag). A second 
factor concerns the scattering efficiency of the grain population. For silicate 
grains with radii between 0.1-1.0 \mum, the ratio of the scattering cross-sections, 
$Q_{sca,F444W}/Q_{sca,V}$ has a roughly constant value of 0.1 \citep{Draine1984}.
Table~\ref{tab:disk} scales the HST limits by these factors and compares them to the
observed F444W values at three of the separations shown in Figure~\ref{fig:azimuth}.
The F444W limits are more stringent than the visible light values, but still do not constitute a significant detection.  Models of the disk emission at multiple wavelengths,   incorporating both  F444W and MIRI data, are discussed in Wolff et al (2025, in prep).

\begin{deluxetable*}{llll}
\tabletypesize{\scriptsize}
\tablewidth{0pt}
\tablecaption{Limits to Disk Emission from HST and JWST\label{tab:disk}
}
\tablehead{
\colhead{Separation }&\colhead{HST Limit$^1$}&\colhead{Extrapolated F444W$^2$}&\colhead{Observed F444W}\\ \colhead{(\arcsec) }&\colhead{MJy Sr$^{-1}$}&\colhead{a=0.1$\sim$1\mum, MJy Sr$^{-1}$}&\colhead{MJy Sr$^{-1}$}}
\startdata
2.0&$<$67&$<$2.4&$0.3\pm0.3$\\
2.5&$<$34&$<$1.2&$<0.1$\\
3.0&$<$21&$<$0.75&$<0.1$\\
\enddata
\tablecomments{$1$From \citet{Krishnanth2024}. $^2$Extrapolated from HST limit taking into reduced stellar brightness and scattering cross-sections for grain radii between 0.1 and 1.0 \mum\ \citep{Draine1984}.}
\end{deluxetable*}

\subsection{ Sources in the Field of View}\label{sec:results_sources}

To retrieve the astrometry and photometry of the point-like sources, we fitted a model to the PSF-subtracted data using an MCMC framework with the \texttt{pyKLIP} package \citep{Wang2015,Foreman-Mackey2013}. Table~\ref{tab:JWSTsources} gives the positions and magnitudes of all sources within the  fields of view  ($\pm$10\arcsec,$\pm$100\arcsec ) of the coronagraph in the subarray and full fields of view.  The majority of these can be tentatively identified as being background stars or galaxies on the basis of their spectral energy distributions, [F210M]-[F444W]$<$0.2, and their presence in other datasets such as HST and ALMA \citep{Wolff2023,Krishnanth2024,Booth2023}.

{In the case of S~1, the photometry and astrometry are estimated within the \texttt{pyKLIP} framework under the assumption that it is a point-source. However, given that this source is located within one of the zones with highly correlated noise, we estimate the error bar for the flux using the standard deviation of the noise at the other two, high noise regions.}

\begin{deluxetable*}{c|c|cccc}
\tabletypesize{\scriptsize}
\tablewidth{0pt}
\tablecaption{JWST Results for \epseri \label{tab:JWSTsources}
}
\tablehead{
\colhead{}& Notes &\colhead{$\Delta$RA$^1$}
& \colhead{$\Delta$Dec$^1$}
&\colhead{F210M}
&\colhead{F444W}\\
\colhead{Source} & & \colhead{(\arcsec)}  
& \colhead{(\arcsec)}& 
\colhead{(uJy)}
&\colhead{(uJy)}
}
\startdata
\multicolumn{6}{l}{}\\
S~1 & Marginal detection &0.28$\pm$0.022&0.86$\pm$0.022&$<$7.9 (5-$\sigma$)&87.3$\pm$10.4 \\
S~2 & Extended &7.75$\pm$0.007&15.83$\pm$0.007&$<$0.17 (5-$\sigma$)& 8.8$\pm$1.4 \\
HST~1 & Point-like; seen by HST &11.30$\pm$0.007&13.45$\pm$0.007&31.79$^{+0.97}_{-0.98}$ &14.0$\pm$1.3 \\
HST~2 & Extended; seen by HST \&~ALMA &-10.54$\pm$0.007&15.47$\pm$0.007&$<$0.18 (5-$\sigma$)& 9.6$\pm$1.0 \\
ALMA~1 & Extended; seen by ALMA &14.69$\pm$0.007&-18.72$\pm$0.007&N/A $^2$&11.9$\pm$1.3 \\ 
\enddata
\tablecomments{$^1$NIRCam astrometry from 2023-Aug-08\\ {$^2$The error bar was estimated with a standard deviation of the noise at highly correlated noise regions. } \\ {$^3$}Outside of the F210M FOV.}
\end{deluxetable*}

The only previously unseen source within 10\arcsec, is Source \#2 shown in Fig.~\ref{fig:NIRCamSources}. It exhibits signs of being extended, so we expect it is most likely a galaxy. This source was also detected in the MIRI dataset under the same program (Wolff et al. (in preparation)).

\section{Discussion\label{sec:discuss}}
\subsection{Planet b}\label{sec:planetb}
As presented in Sec.~\ref{sec:results_inner_region}, a faint, point-source-like feature at a separation of 0.91\arcsec~and PA of 162$^\circ$~is consistently recovered in the subarray dataset. Its flux and position are consistent with the latest model for \epseri~b \citep{Thompson2025}: 0.98$\pm$0.09~\mj. Independent reductions, including a PSF-subtraction using the \texttt{VIP} package~\cite{GomezGonzalez2017}, recover consistent results for this feature. However, since its position is near a bright hexpeckle, where the noise is more intense and non-Gaussian, its significance is hard to evaluate. A new epoch with the NIRCam corongraph, aiming at avoiding the hexpeckles, possibly using the M430R coronagraph mask as opposed to the M335R, could confirm its validity. 
We report its astrometry and photometry under the assumption of it being a point-source in Tab.~\ref{tab:JWSTsources}, labeled S~1.

\begin{figure}
\centering
 \includegraphics[width=0.48\textwidth]{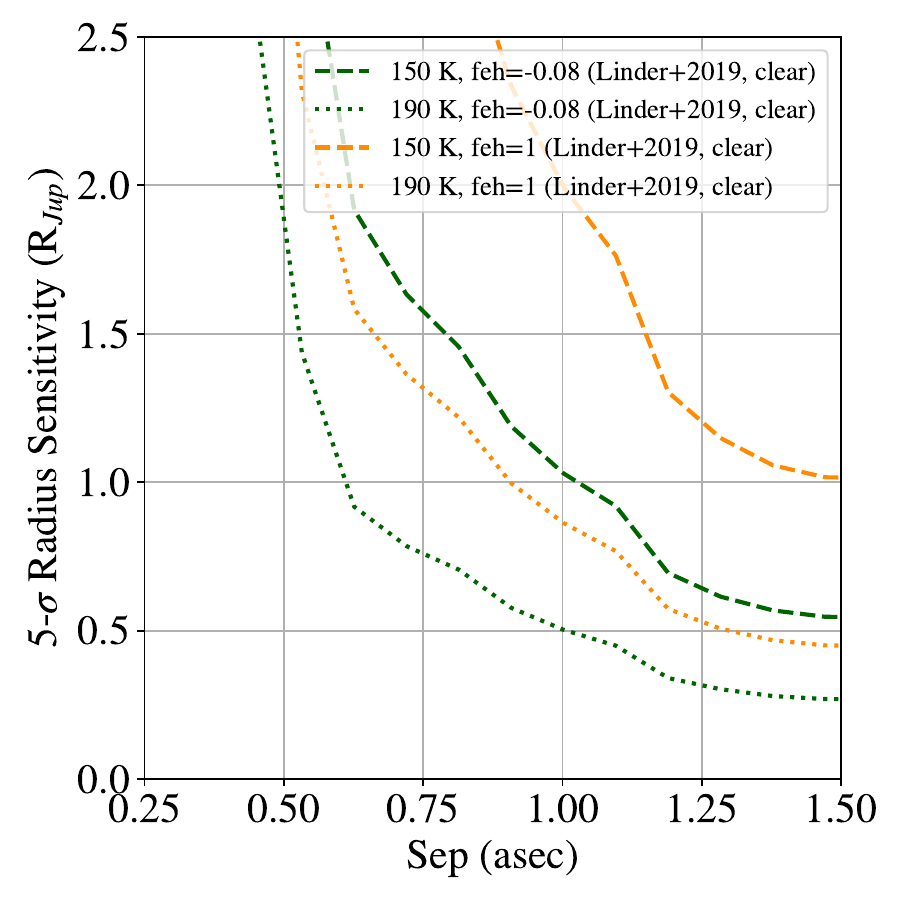} 
\caption{Planet radius limits for a planet of mass=0.98\mj, i.e. the latest mass estimate for \epseri~b \citep{Thompson2025}. These curves are computed with \citet{Linder2019} models and the contrast numbers reported in Fig.~\ref{fig:contrastlimits}. We use the range of temperatures expected for a Jupiter-like planet in the range of \epseri's age. 
\label{fig:radius_sensitivity_b}}
\end{figure}

The contrast and mass sensitivity curves shown in Figs.~\ref{fig:contrastlimits}~and~\ref{fig:mass_sensitivities} are computed excluding the areas near the three bright hexpeckles. For this assumption, the new upper limit is quite strict: at the lower end of the temperature bracket, planet b should have been imaged with high completeness (Fig.~\ref{fig:mass_sensitivities} red contours versus dotted line) assuming a Jupiter-analog. When taking into account different surface gravities and metallicities, the range of planets to which these observations are sensitive to is rather wide. We show these sensitivities in terms of planet radius in  Fig.~\ref{fig:radius_sensitivity_b}.
~To compute these, we assume the latest planet mass estimate of 0.98$\pm$0.09\mj~and place limits on the planet's radius as a function of separation for the observation epoch using the contrast limits at the F444W filter. At a separation of 1\arcsec, the planet radius sensitivities range from 0.5 to 2 R$_{Jup}$ depending on the effective temperature and metallicity assumption.

A potentially more sensitive approach to detecting planet b comes in the form of the JWST/NIRSpec's IFU observations. This mode has been shown to outperform the NIRCam coronagraph in terms of contrast at close separations \citep{Ruffio2024}. A cycle 3 program (P\# 4982, PI: Ruffio) will perform NIRSpec observations of \epseri~targeting planet b, aiming at estimating its effective temperature and atmospheric composition by virtue of the spectroscopic capabilities of NIRSpec's IFU.

\subsection{Three-roll performance}
\begin{figure}[t!]
   \begin{center}
   \begin{tabular}{c} 
   \includegraphics[width=0.43\textwidth,trim={0cm 0cm 0cm 0cm}]{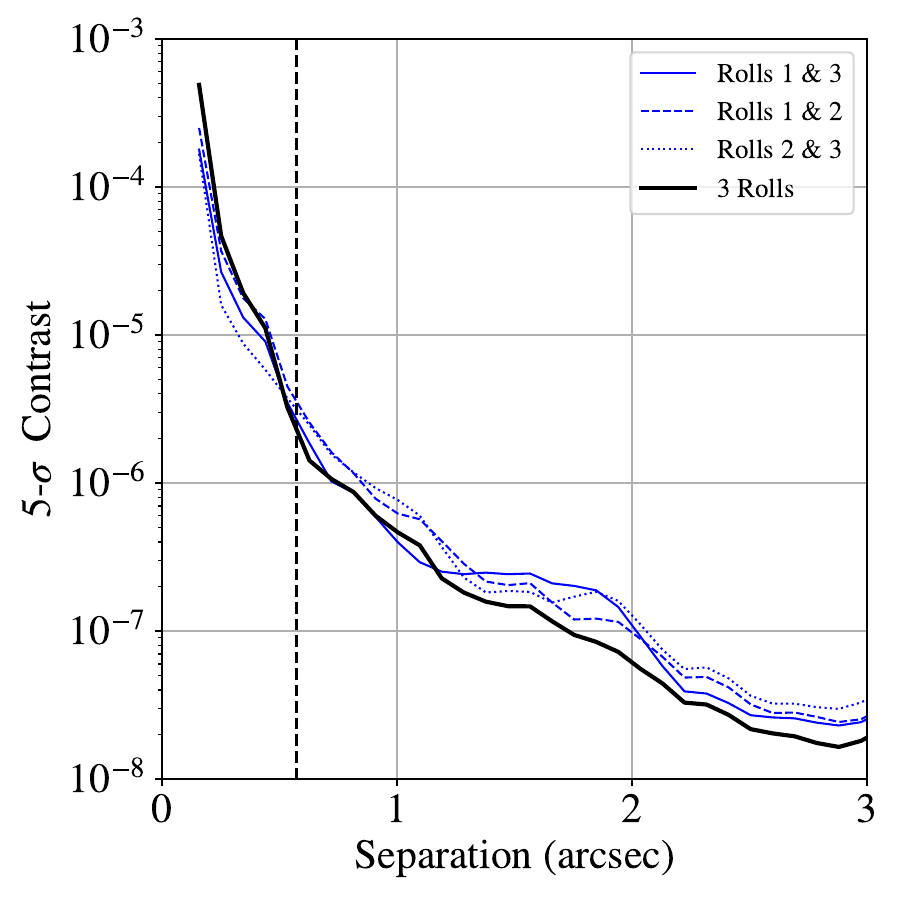}
   \end{tabular}
   \end{center}
   \caption{Contrast curves comparing the performance of the 3-roll strategy (\textit{black}) with respect to pair-wise combinations (\textit{blue}) from the same dataset. These observations of \epseri~were the first to test a 3-roll observation strategy. The 3-roll combination performs generally better beyond the inner working angle of the coronagraph (vertical line). However, the gains are {modest}: the 3-roll curve never outperforms all three 2-roll combinations by a wide margin. At small separations, the small roll angle differences result in self-subtraction, which favors the pair-wise combinations over the 3-roll combination. 
   \label{fig:3rolls}} 
\end{figure}

These observations have been the first to utilize three telescope rolls as opposed to the conventional two-rolls observation. The motivation for this experimental strategy is to obtain increased speckle diversity and enhance the sensitivity to point sources. In Fig.~\ref{fig:3rolls} we show the contrast limits for the 3-roll strategy and how it compares against the independent combination of 2 rolls taken from the same dataset. As expected, the 3-rolls performs better than any 2-roll combination. {However, the gains from this strategy are modest: at small angular separations it yields an average of $\sim$20\% better contrast  with respect tot he 2-roll strategy, but it never perform 2$\times$ better at any separation. }
Within the inner working angle, the effects of self-subtraction, that are accounted on the contrast via the algorithmic throughput (see Sec.~\ref{sec:results_sensitivity}), dominate. {Hence, the 3-roll strategy is not well suited for very small angular separations.}

\subsection{$\delta$~Eridani}
\begin{figure}[t!]
   \begin{center}
   \begin{tabular}{c} 
   \includegraphics[width=0.43\textwidth,trim={0cm 0cm 0cm 0cm}]{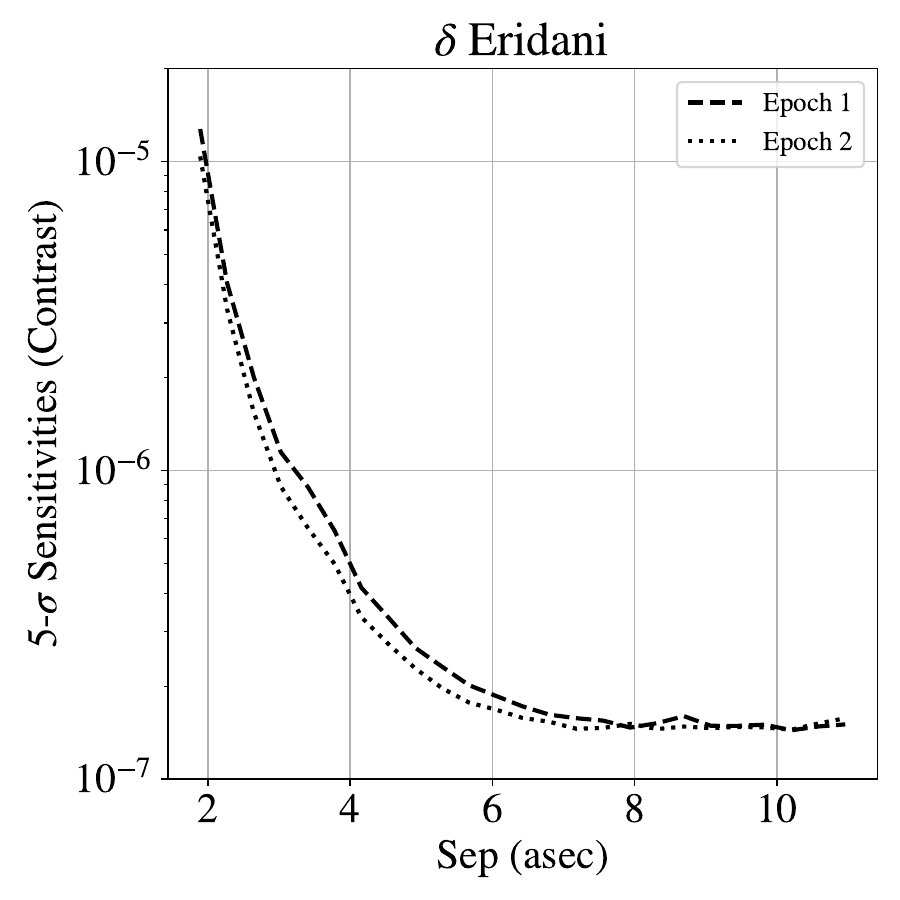}
   \end{tabular}
   \end{center}
   \caption{Contrast curves for the two epochs of observations of \deleri, the reference star for PSF subtraction. To obtain these, we used the \epseri~images as PSF references. The sensitivity difference with respect to \epseri~are due to the lack of two rolls and magnitude difference.
   \label{fig:delEri}} 
\end{figure}

The reference star selected for PSF subtraction was $\delta$~Eridani, which is among the precursor science target list for the Habitable Worlds Observatory (HWO), as identified by the Exoplanet Exploration Program (ExEP) \citep{Mamajek2024}.
Given its status as precursor science for HWO we computed the contrast sensitivities for $\delta$~Eridani, shown in Fig.~\ref{fig:delEri}. These are computed by centering the 5-point-SGD images and co-adding, and performing PSF-subtraction with the \epseri~roll images. To optimize the PSF-subtraction we use a linear combination of the two \epseri~rolls that results in the optimal contrast in a least-squares optimization. As expected, the contrast achieved for the reference frames is not as deep as with the \epseri~data given the brightness difference and lack of rolls. \deleri~is a mature system, of an estimated age of $\sim$6 Gyr \citep{Thevenin2005}, so these contrast limits at 4 $\mu$m are not as relevant as for young systems. {Regarding bound companions, there were no sources detected in the data that shared a common proper motion with \deleri.}

\section{Conclusion}
We presented the JWST/NIRCam observations of \epseri~to search for planets orbiting this system. These observations were performed with the F444W and F210M filters, with two epochs targeting the outer region of the system, and one epoch targeting the RV planet orbiting at $\sim$3.5 AU. 
For the outer region, contrast limits reach 2$\times$\tentoe~beyond 16 AU, which, given current models and the age of \epseri, rule out planets with the mass Saturn or higher at these separations. A source at $\sim$50 AU, S2, is detected, but it shows signs of being extended, making it thus most likely a galaxy.  
Although there was no clear detection of a point-source in the inner region, a persistent blob within 1-$\sigma$ of the expected position of planet b \citep{Thompson2025} consistently appears in the PSF-subtracted data. Its location is within 1~$\lambda/$D of the brightest speckle in the raw data, and the resulting noise is highly non-Gaussian. Thus, we consider this blob as not being statistically significant. Further observations will be needed to test its validity.

\acknowledgements
J.Ll.-S. thanks Virginie Faramaz for a fruitful discussion about \epseri's disk. NIRCam development and use at the University of Arizona is supported through NASA Contract NAS5-02105. Part of this work was carried out at the Jet Propulsion Laboratory, California Institute of Technology, under a contract with the National Aeronautics and Space Administration (80NM0018D0004). The work of A.G., G.R., and S.W.
\ was partially supported by NASA grants NNX13AD82G and 1255094. This material is based upon work supported by the National Science Foundation Graduate Research Fellowship under Grant No.~2139433. The High Performance Computing resources used in this investigation were provided by funding from the JPL Information and Technology Solutions Directorate. We are grateful for support from NASA for the JWST NIRCam project though contract number NAS5-02105 (M. Rieke, University of Arizona, PI).

\copyright 2025. All rights reserved.

\facilities{JWST}

\software{
\texttt{astropy} \citep{astropy2022},
\texttt{jwst} \citep{jwst2022},
\texttt{NIRCoS} \citep{Kammerer2022},
\texttt{pyNRC} \citep{Leisenring2024},
\texttt{pyKLIP} \citep{Wang2015},
\texttt{SpaceKLIP} \citep{Kammerer2022},
\texttt{STPSF} \citep{perrin14},
\texttt{WebbPSF\_ext} \citep{Leisenring2024}
}


\clearpage

\begin{thebibliography}{99}
\expandafter\ifx\csname natexlab\endcsname\relax\def\natexlab#1{#1}\fi
\renewcommand{\bibfont}{\small}
\setlength{\itemsep}{0pt}

\bibitem[{{Amara} \& {Quanz}(2012)}]{Amara2012}
{Amara}, A., \& {Quanz}, S.~P. 2012, \mnras, 427, 948,
 doi:{10.1111/j.1365-2966.2012.21918.x}


 
\bibitem[{{Astropy Collaboration} {et~al.}(2022){Astropy Collaboration},
 {Price-Whelan}, {Lim}, {Earl}, {Starkman}, {Bradley}, {Shupe}, {Patil},
 {Corrales}, {Brasseur}, {N{\"o}the}, {Donath}, {Tollerud}, {Morris},
 {Ginsburg}, {Vaher}, {Weaver}, {Tocknell}, {Jamieson}, {van Kerkwijk},
 {Robitaille}, {Merry}, {Bachetti}, {G{\"u}nther}, \& {Astropy Project Contributors}}]{astropy2022}
{Astropy Collaboration}, {Price-Whelan}, A.~M., {Lim}, P.~L., {et~al.} 2022,
 \apj, 935, 167
 
\bibitem[Backman \& Paresce(1993)]{Backman1993} Backman, D.~E. \& Paresce, F.\ 1993, Protostars and Planets III, 1253

\bibitem[Baraffe et al.(2003)]{Baraffe2003} Baraffe, I., Chabrier, G., Barman, T.~S., et al.\ 2003, \aap, 402, 701. doi:10.1051/0004-6361:20030252

\bibitem[{{Baraffe} {et~al.}(2003){Baraffe}, {Chabrier}, {Barman}, {Allard}, \&
 {Hauschildt}}]{baraffe03}
{Baraffe}, I., {Chabrier}, G., {Barman}, T.~S., {Allard}, F., \& {Hauschildt},
 P.~H. 2003, \aap, 402, 701
 
 \bibitem[Beichman et al (2024)]{Beichman2024} Beichman, C.A., Bryden, G., Llop-Sayson,J et al. submitted.
 
\bibitem[Booth et al.(2017)]{Booth2017} Booth, M., Dent, W.~R.~F., Jord{\'a}n, A., et al.\ 2017, \mnras, 469, 3200. doi:10.1093/mnras/stx1072

\bibitem[Bushouse et~al.\ (2022)]{jwst2022}
Bushouse, H., Eisenhamer, J., Dencheva, N., {et~al.} 2022, JWST Calibration
 Pipeline, Zenodo, doi:10.5281/ZENODO.7038885.
\newblock \url{https://zenodo.org/record/7038885}

\bibitem[Draine \& Lee(1984)]{Draine1984} Draine, B.~T. \& Lee, H.~M.\ 1984, \apj, 285, 89. doi:10.1086/162480


\bibitem[Feng et al.(2023)]{Feng2023} Feng, F., Butler, R.~P., Vogt, S.~S., et al.\ 2023, \mnras, 525, 607. doi:10.1093/mnras/stad2297


\bibitem[Gillett (1986)]{Gillett1986} Gillett F. C., 1986, in Astrophysics and Space Science Library,
Vol. 124, Light on Dark Matter, Israel F. P., ed., pp. 61-69


\bibitem[Hatzes et al.(2000)]{Hatzes2000} Hatzes, A.~P., Cochran, W.~D., McArthur, B., et al.\ 2000, \apjl, 544, L145. doi:10.1086/317319

\bibitem[HCI PSF Reference Stars (2023)]{JDoxColor} {HCI PSF Reference Stars (2023)} \url{https://jwst-docs.stsci.edu/methods-and-roadmaps/jwst-high-contrast-imaging/jwst-high-contrast-imaging-proposal-planning/hci-psf-reference-stars}

\bibitem[Janson et al.(2015)]{Janson2015} Janson, M., Quanz, S.~P., Carson, J.~C., et al.\ 2015, \aap, 574, A120. doi:10.1051/0004-6361/201424944




\bibitem[Johnson et al.(1966)]{Johnson1966} Johnson, H.~L., Mitchell, R.~I., Iriarte, B., et al.\ 1966, Communications of the Lunar and Planetary Laboratory, 4, 99 

\bibitem[{{Kammerer} {et~al.}(2022){Kammerer}, {Girard}, {Carter}, {Perrin},
 {Cooper}, {Thatte}, {Vandal}, {Leisenring}, {Wang}, {Balmer},
 {Sivaramakrishnan}, {Pueyo}, {Ward-Duong}, {Sunnquist}, \& {Adams
 Redai}}]{Kammerer2022}
{Kammerer}, J., {Girard}, J., {Carter}, A.~L., {et~al.} 2022, in Society of
 Photo-Optical Instrumentation Engineers (SPIE) Conference Series, Vol. 12180,
 Space Telescopes and Instrumentation 2022: Optical, Infrared, and Millimeter
 Wave, ed. L.~E. {Coyle}, S.~{Matsuura}, \& M.~D. {Perrin}, 121803N


\bibitem[Krishnanth et al.(2024)]{Krishnanth2024} Krishnanth, P.~M.~S., Douglas, E.~S., Anche, R.~M., et al.\ 2024, \aj, 168, 169. doi:10.3847/1538-3881/ad6efe

\bibitem[Lawson et al.(2022)]{Lawson2022} Lawson, K., Currie, T., Wisniewski, J.~P., et al.\ 2022, \apjl, 935, L25. doi:10.3847/2041-8213/ac853b

\bibitem[Lawson et al.(2024)]{Lawson2024} Lawson, K., Schlieder, J.~E., Leisenring, J.~M., et al.\ 2024, \apjl, 967, L8. doi:10.3847/2041-8213/ad4496

\bibitem[{{Leisenring}(2023)}]{Leisenring2024} {Leisenring}, J. 2024, \apj, in prep


\bibitem[Linder et al.(2019)]{Linder2019} Linder, E.~F., Mordasini, C., Molli{\`e}re, P., et al.\ 2019, \aap, 623, A85. doi:10.1051/0004-6361/201833873


\bibitem[Mamajek \& Hillenbrand(2008)]{Mamajek2008} Mamajek, E.~E. \& Hillenbrand, L.~A.\ 2008, \apj, 687, 1264. doi:10.1086/591785


\bibitem[{{Mawet} {et~al.}(2014){Mawet}, {Milli}, {Wahhaj}, {Pelat}, {Absil},
 {Delacroix}, {Boccaletti}, {Kasper}, {Kenworthy}, {Marois}, {Mennesson}, \&
 {Pueyo}}]{Mawet2014}
{Mawet}, D., {Milli}, J., {Wahhaj}, Z., {et~al.} 2014, \apj, 792, 97,
 doi:{10.1088/0004-637X/792/2/97}

\bibitem[Mawet et al.(2019)]{Mawet2019} Mawet, D., Hirsch, L., Lee, E.~J., et al.\ 2019, \aj, 157, 33. doi:10.3847/1538-3881/aaef8a



\bibitem[{{Perrin} {et~al.}(2014){Perrin}, {Sivaramakrishnan}, {Lajoie},
 {Elliott}, {Pueyo}, {Ravindranath}, \& {Albert}}]{perrin14}
{Perrin}, M.~D., {Sivaramakrishnan}, A., {Lajoie}, C.-P., {et~al.} 2014, in
 Society of Photo-Optical Instrumentation Engineers (SPIE) Conference Series,
 Vol. 9143, Space Telescopes and Instrumentation 2014: Optical, Infrared, and
 Millimeter Wave, ed. J.~{Oschmann}, Jacobus~M., M.~{Clampin}, G.~G. {Fazio},
 \& H.~A. {MacEwen}, 91433X
 
\bibitem[Perrin et al.(2018)]{Perrin2018} Perrin, M.~D., Pueyo, L., Van Gorkom, K., et al.\ 2018, \procspie, 10698, 1069809. doi:10.1117/12.2313552

\bibitem[Rieke et al.(2023)]{Rieke2023} Rieke, M.~J., Kelly, D.~M., Misselt, K., et al.\ 2023, \pasp, 135, 028001. doi:10.1088/1538-3873/acac53

\bibitem[Rosenthal et al.(2021)]{Rosenthal2021} Rosenthal, L.~J., Fulton, B.~J., Hirsch, L.~A., et al.\ 2021, \apjs, 255, 8. doi:10.3847/1538-4365/abe23c

\bibitem[{{Soummer} {et~al.}(2012){Soummer}, {Pueyo}, \&
 {Larkin}}]{Soummer2012}
{Soummer}, R., {Pueyo}, L., \& {Larkin}, J. 2012, \apjl, 755, L28,
 doi:{10.1088/2041-8205/755/2/L28}

 
\bibitem[Spiegel \& Burrows(2012)]{Spiegel2012} Spiegel, D.~S. \& Burrows, A.\ 2012, \apj, 745, 174. doi:10.1088/0004-637X/745/2/174

\bibitem[Su et al.(2017)]{Su2017} Su, K.~Y.~L., De Buizer, J.~M., Rieke, G.~H., et al.\ 2017, \aj, 153, 226. doi:10.3847/1538-3881/aa696b

\bibitem[{{Wang} {et~al.}(2015){Wang}, {Ruffio}, {De Rosa}, {Aguilar}, {Wolff},
 \& {Pueyo}}]{Wang2015}
{Wang}, J.~J., {Ruffio}, J.-B., {De Rosa}, R.~J., {et~al.} 2015, {pyKLIP: PSF
 Subtraction for Exoplanets and Disks}, Astrophysics Source Code Library,
 record ascl:1506.001.
 


\bibitem[Booth et al.(2023)]{Booth2023} Booth, Mark and Pearce, Tim D and Krivov, Alexander V and Wyatt, Mark C and Dent, William R F and Hales, Antonio S and Lestrade, Jean-François and Cruz-Sáenz~de~Miera, Fernando and Faramaz, Virginie C and Löhne, Torsten and Chavez-Dagostino, Miguel, \mnras, 521-4. doi:10.1093/mnras/stad938

\bibitem[Foreman-Mackey et al.(2013)]{Foreman-Mackey2013} Foreman-Mackey, D., Hogg, D.~W., Lang, D., et al.\ 2013, \pasp, 125, 306. doi:10.1086/670067

\bibitem[Llop-Sayson et al.(2021)]{Llop-Sayson2021} Llop-Sayson, J., Wang, J.~J., Ruffio, J.-B., et al.\ 2021, \aj, 162, 181. doi:10.3847/1538-3881/ac134a

\bibitem[Wolff et al.(2023)]{Wolff2023} Wolff, S.~G., G{\'a}sp{\'a}r, A., H. Rieke, G., et al.\ 2023, \aj, 165, 115. doi:10.3847/1538-3881/acac83

\bibitem[Greenbaum et al.(2023)]{Greenbaum2023} Greenbaum, A.~Z., Llop-Sayson, J., Lew, B.~W.P., et al.\ 2023, \apj, 945, 126. doi.org/10.3847/1538-4357/acb68b

\bibitem[Ygouf et al.(2024)]{Ygouf2024} Ygouf, M., Beichman, C.~A., Llop-Sayson, J., et al.\ 2024, \aj, 167, 26. doi:10.3847/1538-3881/ad08c8

\bibitem[Carter et al.(2023)]{Carter2023} Carter, A.~L., Hinkley, S., Kammerer, J., et al.\ 2023, \apjl, 951, L20. doi:10.3847/2041-8213/acd93e

\bibitem[Ruffio et al.(2024)]{Ruffio2024} Ruffio, J.-B., Perrin, M.~D., Hoch, K.~K.~W., et al.\ 2024, \aj, 168, 73. doi:10.3847/1538-3881/ad5281

\bibitem[Mamajek \& Stapelfeldt(2024)]{Mamajek2024} Mamajek, E. \& Stapelfeldt, K.\ 2024, arXiv:2402.12414. doi:10.48550/arXiv.2402.12414

\bibitem[Th{\'e}venin et al.(2005)]{Thevenin2005} Th{\'e}venin, F., Kervella, P., Pichon, B., et al.\ 2005, \aap, 436, 253. doi:10.1051/0004-6361:20042075

\bibitem[Thompson et al.(2025)]{Thompson2025} Thompson, W., Nielsen, E., Ruffio, J.-B., et al.\ 2025, arXiv:2502.20561. doi:10.48550/arXiv.2502.20561

\bibitem[Gomez Gonzalez et al.(2017)]{GomezGonzalez2017} Gomez Gonzalez, C.~A., Wertz, O., Absil, O., et al.\ 2017, \aj, 154, 7. doi:10.3847/1538-3881/aa73d7

\bibitem[Sahlholdt et al.(2019)]{Sahlholdt2019} Sahlholdt, C.~L., Feltzing, S., Lindegren, L., et al.\ 2019, \mnras, 482, 895. doi:10.1093/mnras/sty2732

\bibitem[Backman et al.(2009)]{Backman2009} Backman, D., Marengo, M., Stapelfeldt, K., et al.\ 2009, \apj, 690, 1522. doi:10.1088/0004-637X/690/2/1522

\bibitem[Gray et al.(2003)]{Gray2003} Gray, R.~O., Corbally, C.~J., Garrison, R.~F., et al.\ 2003, \aj, 126, 2048. doi:10.1086/378365

\bibitem[Mu{\~n}oz Bermejo et al.(2013)]{Bermejo2013} Mu{\~n}oz Bermejo, J., Asensio Ramos, A., \& Allende Prieto, C.\ 2013, \aap, 553, A95. doi:10.1051/0004-6361/201220961

\end{thebibliography}
\end{document}